    \documentclass[14pt]{article}
    \usepackage[cp1251]{inputenc}
    \usepackage[russian]{babel}
   \usepackage[russian]{babel}
    \usepackage{graphicx}
    \usepackage{amsmath}
    \usepackage{amsfonts}
    \usepackage{amssymb}
    \usepackage[cp1251]{inputenc}
    \ifx\pdfoutput\undefined
    \usepackage{graphicx}
    \usepackage[usenames]{color}
    \else
    \usepackage[pdftex]{graphicx}
    \fi

\begin{document}
\fontsize{13}{16.8} \selectfont
\renewcommand{\figurename}{Figure}
\renewcommand{\refname}{References}
\renewcommand{\abstractname}{Abstract}
 \author{$ {\mbox{P.I. Fomin}}^{\mbox{a}}  \ , {\mbox{P.A.Nakaznoy}}^{\mbox{b}}
 \ ,{\mbox{S.I.Vilchinskyi}}^{\mbox{b}}
 $
\\ a - Bogolyubov Institute for Theoretical Physics, Kiev, Ukraine
\\ b - Kiev National Taras Schevchenko University, Ukraine
\\ pfomin@bitp.kiev.ua \ , nak@univ.kiev.ua \ , sivil@univ.kiev.ua}

\title{ \bf\LARGE  The Closed Universe Model with Variable
$\Lambda-$term}
\date{}  \maketitle

\begin{abstract}
 We have studied the closed universe
model with the variable cosmological term, which is presented as a
sum of two terms: $\Lambda=\Lambda_0 -k R \ .$ First term
$\Lambda_0$ is a constant and it is describing a sum of quantum
field's zero oscillations. In the geometrical sense it determines
an own constant curvature of the space-time.
 The second term is variable, it is proportional to curvature of the space-time and can be
interpretable as response of vacuum to curvature of the
space-time.
 In the framework of such model we have investigated the time dynamic of
 $\Lambda-$term and try to explain a mechanism which to give rise to a large
value of $\Lambda$ hence leading to the "cosmological constant
problem".
  It was studied models for different dark matter equation of state: $p=0$
  and $p=\varepsilon$.
It was shown that the presence of a variable $\Lambda-$term alters
the evolution of the scale factor $a$ in the very early epoch.  In
the context of this model, a severe decrease of the module of
vacuum energy  density at the primitive stage of Universe's
development is explained.  The numerical results for the value of
the $\Lambda$-term agree with  theoretical predictions for the
early epoch and with recent observation  data.
\end{abstract}

\section{ Introduction}
    Recent observation of Type 1a supernova (based on high red-shift
supernova) \cite{1}-\cite{3} and structure formation models
\cite{4}, \cite{5} indicating an accelerating universe have once
more drawn attention to the possible existence of a small positive
cosmological $\Lambda-$term. As well known, $\Lambda$-term can
produce the field of anti-gravity  that causes the cosmological
expansion to accelerate (see, for example \cite{6}). At a
theoretical level a cosmological term is predicted to arise out of
zero-point quantum vacuum fluctuation of fundamental scalar,
spinor, vector and tensor fields. Although a theoretical predicted
value of the energy density of vacuum usually appears to be much
larger than current observation limit \cite{6}. One method of
resolving the dilemma between a very large cosmological term,
predicted by field theory and an extremely small one, suggested
observation with obvious advantages is  to make the cosmological
term time-depended.

    At the present moment there are several models with variable
$\Lambda$-term and next evolutionary relation for $\Lambda$-term
\cite{7}:
$$\Lambda\sim
t^{-2},\quad \Lambda\sim t^{-\alpha}, \quad \Lambda\sim
A+B\exp{(-\alpha t)}, \quad \Lambda\sim a^{-2},\quad  \Lambda\sim
T^{\alpha},\quad \Lambda\sim a^{-\alpha},$$ $$ \quad \Lambda\sim
\exp{(-\alpha a)}, \quad
 \Lambda\sim H^2, \quad \Lambda\sim
H^2+ Aa^{-\alpha}, \quad \Lambda\sim f(H), \quad
\frac{d\Lambda}{d t} \sim g(\Lambda, H)
$$
where $a$ is the scale factor, $H$ is the Hubble parameter, $T$
is the temperature, $t$ is cosmic time, $A, B, \alpha$ are the
constants. However the most of these models are not necessarily
backed by strong physical arguments and does not lead to adequate
physical results.

 In this work we have  studied  the  closed universe model with
variable cosmological term which is presented as a sum of two
terms. We assume that
\begin{equation}
\Lambda=\Lambda_{0}-k R \label{1.1}
\end{equation}
where $\Lambda_{0}, \quad k$ are constants, $R$ is scalar
curvature of space-time  on the basic next physical arguments.
     As well know Einstein's equations can be derived from the action \cite{8}
$$ A=\frac{c^4}{16 \pi G}\int R \sqrt{-g}d^4x + \frac{1}{c}\int
L_{mat}(\varphi,
\partial_{\nu}\varphi)\sqrt{-g}d^4x, $$ where $L_{mat}$ is the
Lagrangian for matter depending some dynamical variables
generically denoted as $\varphi$. The variation of this action
with respect to these variables $\varphi$ will lead to the
equation of motion for matter in a given background geometry,
while  the variation of the action with respect to the metric
tensor $g_{\alpha\beta}$ leads to the Einstein's equations:
$$R_{\alpha\beta}-\displaystyle
\frac{1}{2}g_{\alpha\beta}R=\frac{8 \pi G}{c^4} T_{\alpha\beta}.$$
As is well known (see for example \cite{7}), the consideration a
new matter action $\displaystyle
\acute{L}_{mat}=L_{mat}-\frac{\Lambda}{\frac{8\pi G}{c^4}}$ where
$\Lambda$ is  constant, does not lead to change of the equation of
motion for the matter under this transformation, but the action
will picks up an extra term proportional to $\Lambda :$
\begin{eqnarray}
\label{action} A=\frac{c^4}{16 \pi G}\int R \sqrt{-g}d^4x +
\frac{1}{c}\int(L_{mat}-\frac{\Lambda c^4}{8 \pi G})\sqrt{-g}d^4x=
\nonumber\\
 =\frac{c^4}{16 \pi G}\int (R - 2\Lambda) \sqrt{-g}d^4x + \frac{1}{c}\int
L_{mat}\sqrt{-g}d^4x
\end{eqnarray}
and Einstein's equations  gets modified -- the "$\Lambda$-term"
appears.  The physical interpretation of $\lambda$-term depending
which of the two forms of the  equation (\ref{action}) for the
action is used. According to choice words somebody, the
cosmological constant has the many faces. The first
interpretation, based on the first line of equation
(\ref{action}), treats $\Lambda$ as the shift in the matter
Lagrangian or, the same, the shift in the matter Hamiltonian.
Such a constant shift in the energy does not affect the dynamics
of matter, while gravity, which couples to the total energy of
the system, picks up an extra contribution in the form of a new
term  in the energy-momentum tensor, and Einstein's equations can
be write in such form:
 $$R_{\alpha \beta}-\frac{1}{2}g_{\alpha \beta}R=\frac{8\pi G}{c^4}
(T_{\alpha \beta}+\frac{c^4}{8\pi G} \Lambda g_{\alpha \beta}).$$
In such interpretation introducing the cosmological term is like
introducing a cosmological vacuum with energy momentum tensor
$T_{\alpha \beta}^{vac}$ given by $T_{\alpha
\beta}^{vac}=\displaystyle \frac{c^4}{8\pi G}\Lambda g_{\alpha
\beta}.$
 In other words -- if the vacuum has energy it
also gravitates. The quantum field theory proves that vacuum has
energy \cite{QFT} and what is more quantum field theory in curved
space-time predicted that the density energy of vacuum is
proportional to scalar curvature in curved space-time
\cite{Starobin}, \cite{Birell}. However, numerical calculations of
the vacuum energy density in this case,
 with account for radiative corrections, lead to divergence  even in the
 single-loop approximation \cite{Birell}.
All these results were obtained in a continuous space-time.
 The situation changes drastically if one recalls that,
 as was shown by P.Fomin  \cite{Fomin}, taking into account self-gravitation makes the physical
 vacuum unstable relative to the "decay" into a set of long-living
 coupled "Planck cells". Every "Planck cell" represents a
 microscopic closed universe with zero total energy.
In other words, taking into account self-gravitation
 of vacuum fluctuations leads to a discrete (lattice) structure
 of the vacuum at Planck sizes. Each such configuration has zero total
 mass, charge, and energy, but a nonzero multipole momentum,
 which leads to a gravitational coupling between "Planck cells",
 and therefore to their curving. A curving at the level of Planck cells
 causes the appearance of a nonzero finite vacuum energy, proportional
 to the scalar curvature (the vacuum energy must be Lorentz covariant!).
Hence, within such an approach one can assume that accounting for
self-gravitation of vacuum fluctuations leads
 to a discrete (quantized) contribution to the scalar curvature
 and the vacuum energy density is proportional to the scalar
 curvature:
$\Lambda \thicksim R.$

This approach, in an interesting way, is in accordance with an
 idea once expressed by A.D.Sakharov \cite{Sakharov}, who suggested to treat
 scalar curvature as the rigidity of space-time.

So, in this interpretation we assume that $$
T_{\alpha\beta}^{vac}= \displaystyle \frac{1}{\varkappa}
(\Lambda_0 -k R) g_{ik} \equiv \lambda_0 -\displaystyle
\frac{k}{\varkappa} R, $$ where $\varkappa =\displaystyle \frac{8
\pi G}{c^4} \ .$ The parameter k makes a sence of space's elastic,
$\Lambda_0$ responds to the sum of quantum field's
zero-fluctuations.

    On the other side the second line in equation (\ref{action}) can be interpreted as
gravitational field, determined  by the Lagrangian of the form
$L_{grav}\sim \displaystyle \frac{1}{G}(R - 2 \Lambda)$ and
interaction with matter described by the Lagrangian $L_{matter}$.
In this interpretation, gravity is described by two parameters -
the Newton's constant $G$ and the cosmological term $\Lambda$. It
is then to modify the left hand side of Einstein's equations and
write
  $$
R_{\alpha\beta}-\frac{1}{2}g_{\alpha\beta}
R-g_{\alpha\beta}\Lambda =\frac{8\pi G}{c^4}T_{\alpha \beta}.$$ It
is important that in this interpretation, the space-time is
treated as curved even in the absence of matter, since the
equation $$ R_{\alpha\beta}-\frac{1}{2}g_{\alpha\beta}
R-g_{\alpha\beta}\Lambda =0 $$ does not admit flat space-time as
solution. Therefore in this case $\Lambda -$term has the
geometrical nature -- it determines an own constant curvature of
the space-time and it is constant -- $\Lambda_0.$

\section{Basic equations}

In fact, it is possible to consider a situation in which both
effects can occur.  If the matter Lagrangian contains energy
densities of vacuum, which is proportional to scalar curvature,
then $\Lambda -$ term can change due to the Richi scalar $R$
changes during the evolution. On the other side, if the
gravitational interaction is described by the Lagrangian of the
form $\displaystyle \frac{1}{G}(R -2\Lambda)$, then there is an
intrinsic cosmological constant as the primordial own curvature of
space-time in nature just as there is a Newtonian gravitational
constant in nature. Hence, within such an approach we will review
the universe model with cosmological term which is presented as a
sum of two terms (\ref{1.1}).

    In framework of such assumption the Einstein's equations will be
have next view:
\begin{equation}
\label{2}
 R_{\alpha\beta} -\frac{1}{2} g_{\alpha\beta}
R-\Lambda_{0} g_{\alpha\beta} =\mathop{\varkappa} T_{\alpha\beta}
- k Rg_{\alpha\beta} \ ,
\end{equation}
where $\varkappa= \displaystyle \frac{8 \pi \gamma}{c^4},$ $
T_{\alpha\beta}=\left( \varepsilon + p \right) \mathop{u_\alpha}
u_\beta -p g_{\alpha\beta}$ -- energy-momentum tensor for a
perfect fluid, $\epsilon$ is the mass-energy density of matter,
$p$ is the pressure and $u^\alpha$ is the four-velocity of matter.

We assume the closed homogeneous  and isotropic universe and used
the Friedmann-Robertson-Walker metric:
\begin{equation}
d s^2=c^2 d t^2 -a^2(t) \left( d \chi^2 +\sin^2 \chi \left( d
\theta^2 + \mathop{\sin^2\theta} d \phi^2 \right) \right) \ .
\label{2.2} \end{equation} In this denotations we have this
Friedmann's equations:
\begin{equation}
\begin{cases}
{\dot{a}}^2+c^2=\frac{\displaystyle 1}{\displaystyle 3} \varkappa
c^2 a^2 (\varepsilon +\varepsilon_V ) \\
\ddot{a}=-\frac{\displaystyle 1}{\displaystyle 6} \varkappa c^2 a
(\varepsilon +3 p-2 \varepsilon_V  ) \ . \end{cases}
\label{2.4}\end{equation} For our model $(\ref{1.1})$ from this
follows:
\begin{equation}
\begin{cases}
{\dot{a}}^2+f c^2=\displaystyle \frac{\varkappa c^2 }{1-4 k}a^2
\left( \varepsilon \left( \displaystyle \frac{1}{3}-k \right) -p k
+\displaystyle \frac{\lambda_0}{3} \right) \\
\ddot{a}=\displaystyle -\frac{\varkappa c^2}{1-4 k} a \left(
\varepsilon \left( \displaystyle \frac{1}{6}-k \right) +p \left(
\displaystyle \frac{1}{2}-k \right) -\displaystyle
\frac{\lambda_0}{3} \right) \ , \\ p=\nu \varepsilon \ ,
\end{cases} \label{Fr}\end{equation} where $\nu_i$ is determined
the equation state selection: $\nu=0$ for zero pressure,
$\displaystyle \nu=\frac{1}{3}$
 for pure radiation and $\nu=1$ for limited hard state of a
 matter.

 From Fridman's equations (\ref{Fr})
following the principle of energy density changing
\begin{equation}
\label{4} \varepsilon a^\sigma=Const  , \qquad \sigma=\frac{3(1-4
k)(\nu+1)}{1-3 k(\nu+1)}.
 \end{equation}
According to  different equation of state $\sigma$ is also
different, therefore:
\begin{equation}
\begin{cases}
\mbox{dust:}\, \nu=0 \Longrightarrow \varepsilon a^{
\frac{3(1-4 k)}{1-3 k}}=\varepsilon_0 a_0^{
\frac{3(1-4 k)}{1-3 k}}  \\ \mbox{pure radiation:}\,
\nu=\displaystyle \frac{1}{3} \Longrightarrow \varepsilon
a^4=\varepsilon_0 a_0^4  \\ \mbox{limited hard state of a matter:}
\, \nu=1 \Longrightarrow \varepsilon a^{ \frac{6(1-4 k)}{1-6
k}}=\varepsilon_0 a_0^{ \frac{6(1-4 k)}{1-6 k}},
\end{cases}
\label{5}
\end{equation}
where $\varepsilon_0 $ -- modern values of the matter density for
different kind of  matter, $a_0=\displaystyle \frac{c}{H_0}$ --
modern value of the Hubble radius, $H_0=65 \pm 15
\frac{\mbox{km}}{\mbox{sec}  \, \mbox{Mps}} - \mbox{ modern value
of the Hubble parameter.}$ For next evaluations we are going to
consider, that $$\varepsilon=\sum_i \varepsilon_i \ , p=\sum_i
\nu_i \varepsilon_i \ . $$ First of all, lets research a
concordance of our model with the data of last observations
\cite{Chernin}:
\begin{center}
\begin{tabular}{|l|l|}

  \hline
   type of matter & relative density $ \Omega_{i 0} =\displaystyle \frac{\rho_{i 0}}{\rho_c} $\\
  \hline
   vacuum  & $ \Omega_{\Lambda 0}= 0.7\pm 0.1 $\\
  \hline
   dark matter & $ \Omega_{D 0} = 0.3 \pm 0.1 $ \\
 \hline
   baryon matter & $ \Omega_{B 0} = 0.02 \pm 0.01 $ \\
 \hline
   radiation & $ \Omega_{R 0} = 0.8 \cdot 10^{-4} \alpha \ , \alpha \backsim 1\div 10 $ \\
  \hline

\end{tabular}\end{center} There
\begin{equation}
\Omega =\frac{\rho}{{\rho}_c} \ , \end{equation}
\begin{equation}
{\rho}_c = \frac{3 {H_0}^2}{8 \pi G} = (0.6 \pm 0.1) \times
{10}^{-29} \frac{\mbox{g}}{{\mbox{sm}}^{3}}
\end{equation}
\begin{equation}
H_0=65 \pm 15 \frac{\mbox{km}}{\mbox{s $\cdot$ Mpk}} \mbox{ }
\mbox{- Hubble's constant} \ .\label{1.2}\end{equation}

Certainly, use this data for our model is an approximation. But,
it is a good approximation, because now we have very small ${R
\sim \Lambda} \ .$

Lets equate any appropriate equations of the systems $(\ref{2.4})$
and $(\ref{Fr})$. As a result we are taken:
\begin{equation}
\Omega_{0}=\Omega_{\Lambda 0}-k (4 \Omega_{\Lambda 0}+\Omega_{D 0}
(1-3 \nu_D)+\Omega_{B 0}) \ , \label{2.6} \end{equation} where
$\nu_D$ is the parameter of dark matter state's equation. We use
$p=0$ for a barion matter and  $p=\displaystyle
\frac{\varepsilon}{3}$ for a radiation. We don't know a nature of
the dark matter, therefore use for it this general state's
equation.

Under $z \cong 0.7$, allegedly, we have zero acceleration of the
Universe \cite{Chernin}. Hence we can obtain unknown $k$. For this
equate acceleration under this $z$ to zero, the principle of
energy density changing is taking into account.  As a result we
are taken this transcendental equation for $k$:
\begin{equation}
k=\displaystyle \frac{\Omega_D (1+3 \nu_D)+\Omega_B+2 \Omega_R-2
\Omega_{\Lambda 0}}{2 (3 \Omega_D (1+\nu_D)+3 \Omega_B+4
\Omega_R-4 \Omega_{\Lambda 0}-\Omega_{D 0} (1-3 \nu_D)-\Omega_{B
0})} \label{2.61} \ , \end{equation} where ${\Omega_i=\Omega_{i 0}
(1+z)^{\sigma_i}} \ .$

For numerical calculations we use two possible state's equations
of the dark matter: $\nu_D=0$, i.e. dust and $\nu_D=1$, which is
fiting with limit hard state of matter $\cite{Zeldovich}$. For
this two cases from $(\ref{2.61})$ follows:
\begin{equation}
\begin{cases}
\nu_D=0 \Longrightarrow k=0.03 \ , \Omega_0=0.6 \\ \nu_D=1
\Longrightarrow k=0.03 \ , \Omega_0=0.63 \ . \label{2.62}
\end{cases} \end{equation} Comment, that we have equal values $k$
within the bounds of modern data.

For the next calculations we use $a_0=\displaystyle \frac{c}{H_0}
\ , $ i.e. use hubble's radius. The principle of energy density
changing will be considered to be true for every kind of matter.
Then we can evaluate $\Lambda$-term as function from $a$:
\begin{equation}
\Lambda=\displaystyle \frac{3 \Omega_0}{(1-4 k)
{a_0}^2}+\displaystyle \frac{3 k}{1-4 k} \left((1-3 \nu_D)
\Omega_{D 0} \displaystyle
\frac{{a_0}^{\sigma_D-2}}{a^{\sigma_D}}+\Omega_{B 0} \displaystyle
\frac{{a_0}^{\sigma_B-2}}{a^{\sigma_B}} \right) \label{2.8} \ .
\end{equation} For the concerned cases:
\begin{equation}
\begin{cases}
\nu_D=0 \Longrightarrow \Lambda=\displaystyle \frac{3
\Omega_0}{(1-4 k) {a_0}^2}+\displaystyle \frac{3 k}{1-4 k}
(\Omega_D+\Omega_B) \displaystyle
\frac{{a_0}^{\sigma_0-2}}{a^{\sigma_0}} \ ,\sigma_0=\displaystyle
\frac{3 (1-4 k)}{1-3 k} \\ \nu_D=1 \Longrightarrow
\Lambda=\displaystyle \frac{3 \Omega_0}{(1-4 k)
{a_0}^2}+\displaystyle \frac{3 k}{1-4 k} \left( \Omega_B
\displaystyle \frac{{a_0}^{\sigma_0-2}}{a^{\sigma_0}}-2 \Omega_D
\displaystyle \frac{{a_0}^{\sigma_1-2}}{a^{\sigma_1}} \right) \
,\sigma_1=\displaystyle \frac{6 (1-4 k)}{1-6 k} \ . \label{2.8_1}
\end{cases} \end{equation}

 From $(\ref{2.8_1})$ it follows, that
$\Lambda$-term is positive, when ${0<k<\displaystyle \frac{1}{3}}$
for the barion model of a dark matter and ${k<0}$ for limited hard
state. This is means, that limited hard state of a matter is
requires with value $k=0.03$ for a negative $\Lambda$-term in the
development of Universe.

For the purposes of illustrations $(\ref{2.8_1})$ we make next two
plots, which describing the dependence of the non-dimensional
$\Lambda$-term ${\Omega_\Lambda=\displaystyle
\frac{\Lambda}{\varkappa \rho_c c^2}}$ from the ${x=\displaystyle
\frac{a}{a_0}}$.
\begin{figure}[t]
\begin{tabular}{c}
\includegraphics[width=120mm, angle=-0]{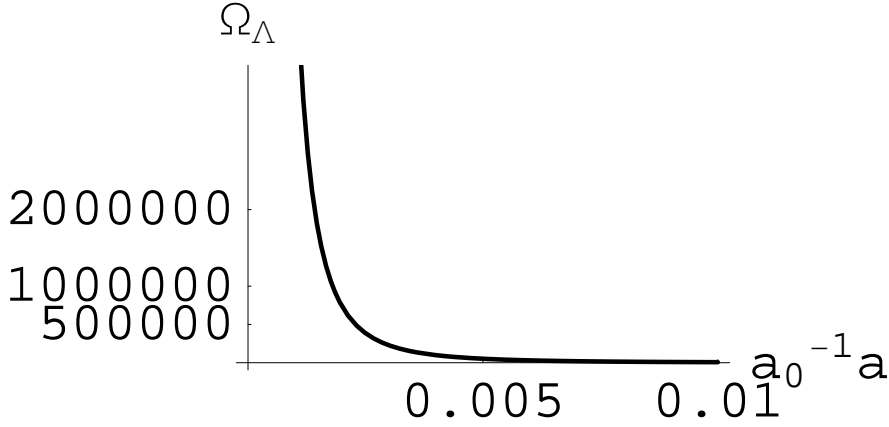}
\end{tabular}
\caption{The dependence of the non-dimensional $\Lambda$-term
${\Omega_\Lambda=\displaystyle \frac{\Lambda}{\varkappa \rho_c
c^2}}$ from the ${x=\displaystyle \frac{a}{a_0}}$ for the barion
model of a dark model, $k=0.03 \ , \Omega_0=0.6 \ .$}
\end{figure}
\begin{figure}[t]
\begin{tabular}{c}
\includegraphics[width=120mm, angle=-0]{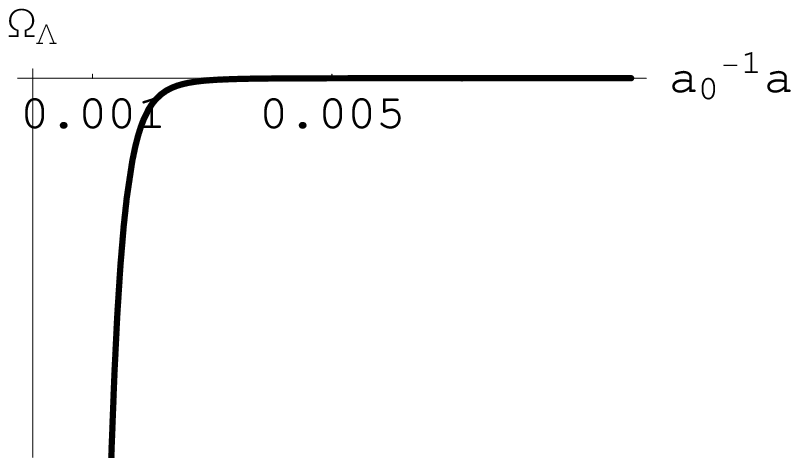}
\end{tabular}
\caption{ The dependence of the non-dimensional $\Lambda$-term
${\Omega_\Lambda=\displaystyle \frac{\Lambda}{\varkappa \rho_c
c^2}}$ from the ${x=\displaystyle \frac{a}{a_0}}$ for the limited
hard model of a dark model $k=0.03 \ , \Omega_0=0.63 \ .$}
\end{figure}

With the help of the principle of energy density changing
$(\ref{4})$ we can write the integral for $a(t)$. In
non-dimeonsional variables $x=\displaystyle \frac{a}{a_0}$ and
$t=\tau H_0$ it has such form:
\begin{equation}
\tau=\int_0^x \frac{d x'}{\sqrt{\displaystyle \frac{1}{1-4 k}
\sum_i (1-3 k(\nu_i+1)) \Omega_{i 0}
{x'}^{(2-\sigma_i)}+\displaystyle \frac{\Omega_0}{1-4 k}
{x'}^2-1}} \ .\label{2.9} \end{equation}

For the two considered models of a dark matter:
\begin{equation}
\begin{cases}
\nu_D=0 \Longrightarrow \\ \tau=\displaystyle \int_0^x \frac{d
x'}{\sqrt{\displaystyle \frac{1-3 k}{1-4 k} (\Omega_{D
0}+\Omega_{B 0}) \mbox{ } {x'}^{\mbox{ } \displaystyle -\frac{1-6
k}{1-3 k}} + \mbox{ } \displaystyle \frac{\Omega_{R 0}
}{{x'}^2}+\displaystyle \frac{\Omega_0}{1-4 k} {x'}^2-1}}
\\ \nu_D=1 \Longrightarrow \\ \tau=\displaystyle \int_0^x \frac{d
x'}{\sqrt{\displaystyle \frac{1-6 k}{1-4 k} \Omega_{D 0} {x'}^{
\displaystyle -\frac{4 (1-3 k)}{1-6 k}}+\frac{1-3 k}{1-4 k}
\Omega_{B 0} \mbox{ } {x'}^{\mbox{ } \displaystyle -\frac{1-6
k}{1-3 k}} + \mbox{ } \displaystyle \frac{\Omega_{R 0}
}{{x'}^2}+\displaystyle \frac{\Omega_0}{1-4 k} {x'}^2-1}}
\label{2.9_1} \ . \end{cases}
\end{equation}

The results of numerical calculations of this integrals presents
at the following plot.

\begin{figure}[t]
\begin{tabular}{c}
\includegraphics[width=130mm, angle=0]{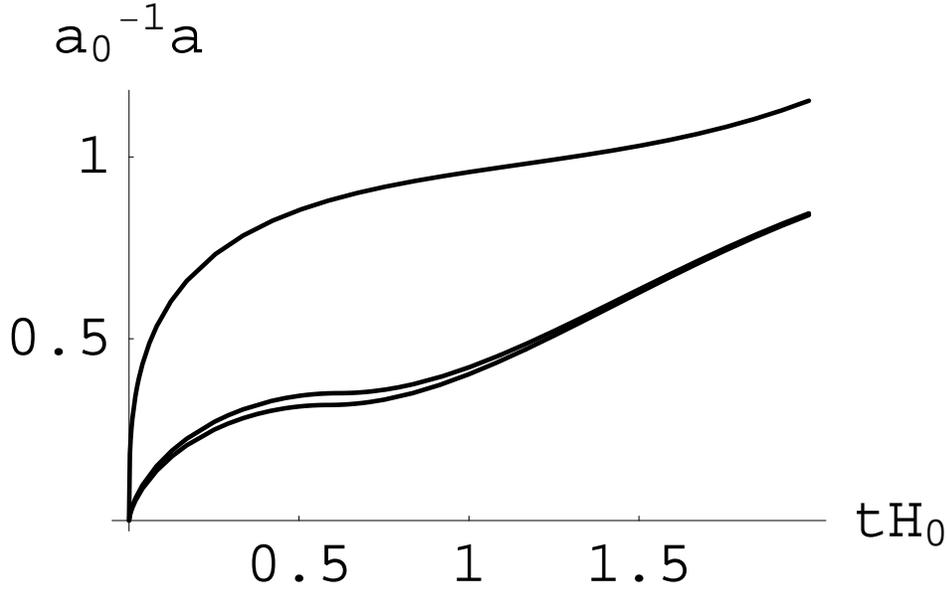}
\end{tabular}
\caption{The dependence of the non-dimensional variable
$x=\displaystyle \frac{a}{a_0}$ from the non-dimensional time
$\tau=t H_0 \ .$ The lower curve is responding to the barion model
of a dark matter, $k=0.03 \ , \Omega_0=0.6 \ .$ The high is
responding to the limited hard model of a dark matter, $k=0.03 \ ,
\Omega_0=0.63 \ .$ The medium is responding to the standard
cosmological model with a flat universe and constant cosmological
term.}
\end{figure}
Under little $x$ this integrals may be calculate in analytic form
and solve to $x$:
\begin{equation}
\begin{cases}
\nu_D=0 \Longrightarrow x(\tau)=\displaystyle (\Omega_{R
0})^{\displaystyle \frac{1}{4}} \sqrt{2 \tau} \\ \nu_D=1
\Longrightarrow x(\tau)=[\displaystyle \sqrt{\frac{1-4k}{1-6k}
\Omega_{D 0}} \mbox{ } 3 \tau ]^{\displaystyle \frac{1-6 k}{3 (1-4
k)}} \ . \label{2.10} \end{cases} \end{equation}

The results of calculations are reducing to next tables. The table
№1 is responding to the barion model of a dark matter (${k=0.03} \
, {\Omega_0=0.6)} \ ,$ №2 - limited hard (${k=0.03} \ ,
{\Omega_0=0.63)} \ .$ Calculations $a(t)$ from $t=10^{-43}
\mbox{с}$ to $t=10^{-10} \mbox{с}$ maked with the $(\ref{2.10})$,
for another points we use numerical calculations $(\ref{2.9_1})$.

Notice, that use of our model formula for the $\Lambda$ - term is
unjustified, when we are speaking about the planck time.
Therefore, values of the cosmological term and dark matter's
densities in this times don't have pretensions of irrefragable
coincidences with anticipated values. This results is used only
for illustration of abrupt slump $\Lambda$-term in an Universe's
first seconds. Nevertheless limited hard model of a dark matter
assures the more steep inflation stage. Moreover we have for the
planck time good coincidence with a planck matter density and
expected value $\Lambda$-term$ \mbox{ }\Lambda\sim{10}^{63}
\displaystyle \frac{1}{{\mbox{sm}}^2}$ until the phase
transformation $SU(5)\longrightarrow SU(3)\times SU(2)\times U(1)
\ ,$ which happened at $t={10}^{-35} \mbox{s}$ and $T={10}^{15}
\mbox{Gev} \ .$ This features makes this model enough attractive.
Notice, that in this model cosmological term is negative until
${t\cong 5 \cdot 10^{16} \mbox{s}}$ ${(a(t)\cong 5.75 \cdot
10^{27} \mbox{sm})} \ .$ Anyway this model expects an additional
researches. There notice, that limited hard state of a matter is
complying with a system of particles with spin $1$. The QFT's
consideration of this system is showing a many interesting
features \cite{Zeldovich}. Particularly, the particles in this
model repels, but antiparticles gravitates. The particles which
have a spin $1$ is gluons and intermediate vector bosons. However,
this particles are constrained and, consequently, haven't required
for the limited hard model of a dark matter features.

    Table 1. {\it Barion model of a dark model, $k=0.03 \ ,
\Omega_0=0.6 $}

\vspace*{1.2cm}\noindent\begin{tabular}{|c|c|c|c|c}

\hline  t (s) & $a(t),$ sm
  &$\Lambda=\Lambda_0-kR$, sm$^{-2}$ &$\rho_D \ ,
  \frac{\mbox{g}}{{\mbox{sm}}^3}$
  \\ \hline  $10^{-43}$ & $  9 \cdot 10^{-4}$ & $4 \cdot 10^{32}$
  & $2 \cdot 10^{60}$
\\ \hline  $10^{-35}$ & $ 9 $ & $9 \cdot 10^{20}$ & $5 \cdot
10^{48}$
 \\ \hline $10^{-10}$ & $ 3 \cdot 10^{13}$ & $4 \cdot 10^{-16}$ &
 $2 \cdot 10^{12}$
\\ \hline  $10^{-5}$ & $8 \cdot 10^{15}$ & $4 \cdot 10^{-23}$ & $2
\cdot 10^5 $
\\ \hline  $10^{-2}$ & $2 \cdot 10^{17}$ & $4 \cdot 10^{-27}$ &
$2$
\\ \hline  $10^{-1}$ & $5 \cdot 10^{17}$ & $3 \cdot 10^{-28}$ &
$1$
\\ \hline  $10^{0}$ & $2 \cdot 10^{18}$ & $4 \cdot 10^{-30}$ & $2
\cdot 10^{-2}$
\\ \hline  $10^{2}$ & $2 \cdot 10^{19}$ & $6 \cdot 10^{-33}$ & $3
\cdot 10^{-5}$
\\ \hline  $10^{7}$ & $6 \cdot 10^{21}$ & $4 \cdot 10^{-40}$ & $2
\cdot 10^{-12}$
\\ \hline  $10^{12}$ & $2 \cdot 10^{24}$ & $2 \cdot 10^{-47}$ &
$10^{-19}$
\\ \hline  $10^{15}$ & $10^{26}$ & $2 \cdot 10^{-52}$ & $10^{-24}$
\\ \hline  $10^{16}$ & $6 \cdot 10^{26}$ & $10^{-54}$ & $6 \cdot
10^{-27}$
\\ \hline  $10^{17}$ & $2 \cdot 10^{27}$ & $6 \cdot 10^{-56}$ & $2
\cdot 10^{-28}$
\\ \hline  $10^{18}$ & $9 \cdot 10^{27}$ & $2 \cdot 10^{-56}$ & $2
\cdot 10^{-30}$
 \\ \hline
\end{tabular}

Table 2. {\it Limited hard model of a dark matter, $k=0.03 \ ,
\Omega_0=0.63$}

\vspace*{1.5cm}\noindent\begin{tabular}{|c|c|c|c|c}

\hline  t (s) & $a(t),$ sm
  &$\Lambda=\Lambda_0-kR$, sm$^{-2}$ &$\rho_D \ ,
  \frac{\mbox{g}}{{\mbox{sm}}^3}$

  \\ \hline  $10^{-43}$ & $2\cdot 10^{9}$ & $-2 \cdot 10^{63}$ &
  $5 \cdot 10^{90}$
\\ \hline  $10^{-35}$ & $ 7 \cdot 10^{11}$ & $ -6 \cdot 10^{46}$ &
$2 \cdot 10^{74} $
 \\ \hline $ 10^{-10}$ & $ 4 \cdot 10^{19}$ & $-7 \cdot 10^{-4}$ &
 $ 2 \cdot 10^{24}$
\\ \hline  $10^{-5}$ & $10^{21}$ & $-7 \cdot 10^{-13}$ & $2 \cdot
10^{15}$
\\ \hline  $10^{-2}$ & $9 \cdot 10^{21}$ & $-5 \cdot 10^{-19}$ &
$2 \cdot 10^9$
\\ \hline  $10^{-1}$ & $2 \cdot 10^{22}$ & $-3 \cdot 10^{-21}$ &
$9 \cdot 10^6$
\\ \hline  $10^{0}$ & $4 \cdot 10^{22}$ & $-4 \cdot 10^{-23}$ &
$10^5$
\\ \hline  $10^{2}$ & $ 2 \cdot 10^{23}$ & $-10^{-27}$ & $3$
\\ \hline  $10^{7}$ & $6 \cdot 10^{24}$ & $-3 \cdot 10^{-37}$ &
$10^{-9}$
\\ \hline  $10^{12}$ & $2 \cdot 10^{26}$ & $-5 \cdot 10^{-47}$ &
$2 \cdot 10^{-19}$
\\ \hline  $10^{15}$ & $2 \cdot 10^{27}$ & $-2 \cdot 10^{-53}$ &
$6 \cdot 10^{-26}$
\\ \hline  $10^{16}$ & $4 \cdot 10^{27}$ & $-2 \cdot 10^{-55}$ &
$7 \cdot 10^{-28}$
\\ \hline  $10^{17}$ & $7 \cdot 10^{27}$ & $2 \cdot 10^{-56}$ & $2
\cdot 10^{-29}$
\\ \hline  $10^{18}$ & $10^{28}$ & $2 \cdot 10^{-56}$ & $2 \cdot
10^{-30}$
 \\ \hline
\end{tabular}

In conclusion we are going to consider the newtonian limit for our
model. From $\cite{8}$ we have:
\begin{equation}
R_0^0=\displaystyle \frac{1}{c^2} \Delta \varphi \ ,
T_0^0=T=\varepsilon=\rho c^2 \ , \end{equation} where $\varphi$ is
the gravity potential. As result, we take this equation:

\begin{equation}
\Delta\varphi=4 \pi G \frac{1-6 k}{1-4 k} \rho-\frac{c^2
\Lambda_0}{1-4 k} \ . \end{equation} The item, which is
proportional to $\Lambda_0$, evidently, incidentally for the
newtonian limit. There we have, however, an important result:
renormalization of the gravitational constant. It, as coefficient
in the action for the gravitational field must be positive. From
this we find one more condition $k<\displaystyle \frac{1}{6}$ or
$k>\displaystyle \frac{1}{4} \ .$ Our value $k=0.03$ belongs to
this interval and, moreover, in view of the fact his smallness,
makes the difference between cosmological and newtonian values $G$
very small : $G_{\mbox{Newt}}\approx 0.93 G$. This difference is
incidentally too, because another cosmological parameters have
larger errors, for example, Hubble's constant $(\ref{1.2})$.
Therefore we don't make a difference between renormalized and
notrenormalized values of $G$.

\section{ Short conclusions}

    1). A model of a closed universe with a time-variant vacuum
energy density is proposed.

    2). It was shown that the presence of a variable $\Lambda-$term
alters the evolution of the scale factor $a$ in the very early
epoch.

    3). In the context of this model, a severe decrease of the vacuum
energy  density at the primitive stage of Universe's development
is explained.  The numerical results for the value of the
$\Lambda$-term agree with  theoretical predictions for the early
epoch and with recent observation  data.

    4) Such explanations are speaking about more attractive the limited
hard model of a dark matter, that requires additional its in-depth
study.

    5) As is shown, variable cosmological term is reducing to
    renormalization of the gravitational constant in the newtonian
    limit.

\end{document}